\documentclass[prb,twocolumn,superscriptaddress,floatfix]{revtex4}
\usepackage[]{graphicx}
\usepackage[]{amsmath}
\usepackage[]{bm}  

\renewcommand{\vec}[1]{\mbox{\boldmath$#1$}}
\newcommand{\crs}{\mbox{\boldmath$\times$}}

\newcommand{\dif}{\mathrm{d}}

\newcommand{\be}{\begin{equation}}
\newcommand{\ee}{\end{equation}}

\newcommand{\F}{\mathcal{F}}

\begin{document}

\bibliographystyle{apsrev}


\title{Tunable magnetization damping in transition metal ternary alloys}

\author{S.~Ingvarsson}
\altaffiliation{Present address: Icelandic Technological Institute, Keldnaholt, IS-112 Reykjavik, Iceland.}
\affiliation{Brown University, Providence, RI 02912}
\affiliation{IBM Research Division, T.~J.~Watson Research Center, Yorktown Heights, NY 10598}
\author{Gang Xiao}
\affiliation{Brown University, Providence, RI 02912}

\author{S.~S.~P.~Parkin}
\affiliation{IBM Research Division, Almaden Research Center, Almaden, Ca 95120}

\author{R.~H.~Koch}
\affiliation{IBM Research Division, T.~J.~Watson Research Center, Yorktown Heights, NY 10598}

\date{\today}

\begin{abstract}
We show that magnetization damping in Permalloy, Ni$_{80}$Fe$_{20}$ (``Py''),
can be enhanced sufficiently to reduce post-switching magnetization precession
to an acceptable level by alloying with the transition metal osmium (Os).  The
damping increases monotonically upon raising the Os-concentration in Py, at
least up to 9\% of Os.  Other effects of alloying with Os are suppression of
magnetization and enhancement of in-plane anisotropy.  Magnetization damping also
increases significantly upon alloying with the five other transition metals
included in this study (4d-elements: Nb, Ru, Rh; 5d-elements: Ta, Pt) but never
as strongly as with Os.
\end{abstract}

\maketitle

Magnetic Tunnel Junctions (MTJ) are presently under intense development for use in magnetic random access memory (MRAM)\cite{ParkinJAP99,Reohri3ecirc02}.  In a slightly modified configuration from the MRAM they can be made extremely sensitive to changes in magnetic field\cite{LiuJAP03,TondraJAP00}.  Such microscopic magnetic field sensors are employed in magnetic recording read heads, readily resolving magnetic bits smaller than 0.1~$\mu$m.  Alternatively MTJ sensors can be scanned along surfaces making a magnetic microscope with great spatial resolution\cite{SchragAPL03,DienyJMMM94}.  The functionality of these devices is based on a change in resistance of the MTJ in response to a change in relative magnetic orientation of a ``free layer'' with respect to a magnetically ``pinned layer''.  These layers are separated (and exchange decoupled) by an insulating oxide tunnel barrier.  Essential to the high-speed functionality of these devices is the swift response of the free layer to the applied field.  It must switch between equilibrium positions defined by the magnetic field conditions, with minimal ``unnecessary'' magnetization motion.  Excessive magnetization precession during the device's relaxation toward equilibrium can cause serious delays.  In the MRAM case the unwanted precession shows up as oscillations in the resistance \emph{after} switching has taken place, often referred to as magnetic ``ringing''\cite{KochPRL98}.  In some cases this even leads to uncertainty of the final state of the device for a given applied field.  

\begin{figure}[htb]
    \begin{center}
	\includegraphics[width=8.5cm]{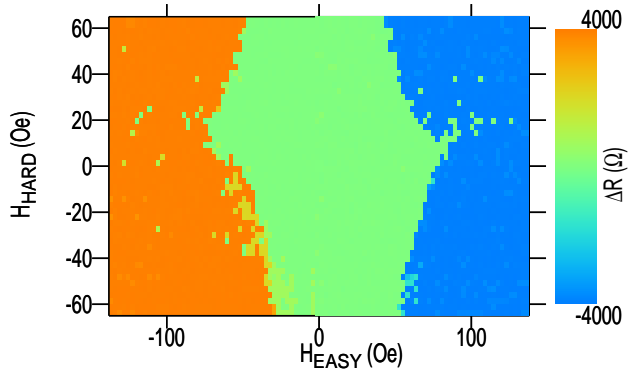}
    \end{center}
    \caption{Ingvarsson \emph{et al.}:  Critical switching curve for a hexagon-shaped MRAM cell with dimensions 0.28~$\mu$m by 0.84~$\mu$m.  The initial state is indicated by a green color, regardless of whether the configuration is parallel or antiparallel.  The resistance scale depicts resistance \emph{relative to the initial state}.  Thus going from right to left (left to right) in the figure the resistance increases (decreases) as the configuration switches at the left (right) green boundary.  The switching was carried out with magnetic field pulses of varying strength, 4~ns in duration with the rise of the easy-axis pulse delayed by 2~ns with respect to the rising edge of the hard-axis pulse.}
    \label{fig:astroid}
\end{figure}

An example of these adverse affects of magnetization precession is displayed in Fig.~\ref{fig:astroid}, measured on a hexagonally shaped MRAM cell with in-plane extremal dimensions of 0.28~$\mu$m and 0.84~$\mu$m.  The MRAM has two equilibrium positions, with parallel (low resistance) and antiparallel magnetization (high resistance).  The figure displays a critical switching curve, extracted from two experiments, one switching the MRAM from parallel to antiparallel configuration (going from right to left in the figure, determining the left boundary of the green area, where the free layer is reversed), and the other switching from antiparallel to parallel (left to right, resolving the right boundary of the green area).  In both experiments the initial state is indicated by a green color, regardless of whether the configuration is parallel or antiparallel.  The resistance scale depicts resistance \emph{relative to the initial state}.  The switching was carried out with magnetic field pulses of varying strength, 4~ns in duration with the rise of the easy-axis pulse delayed by 2~ns with respect to the rising edge of the hard-axis pulse. These experiments correspond to measuring the Stoner-Wolfarth astroid by applying short magnetic field pulses.  Two features in particular set the critical switching curves measured with short field pulses apart from those measured with slowly varying fields.  First are the wide openings at the top and bottom around zero easy-axis field.  Second are the individual unsuccessful switching attempts where the MRAM refused to switch even though the applied fields lie far beyond the continuous part of the curve.  We refer to these anomalous events as ``freckles''.  Freckles are stochastic in the sense that they are not entirely reproducible, but do appear in the same general area of two dimensional field space.  They disappear upon lengthening the field pulses, but appear for different combinations of nanosecond scale pulse lengths with different delay times between hard- and easy-axis pulses.  They are believed to be caused by precession of the magnetic moment after the free layer magnetization has been switched.  If the precession amplitude is large a small perturbation can force the MRAM from the switched state back to the original state.

Efforts to reduce post-switching precession in thin films suitable for MRAM elements include shaping of the magnetic field pulses employed to switch the devices and using so-called precessional switching\cite{CrawfordAPL00, GerritsNAT02, SchumacherPRL03b}.  In precessional switching a perpendicular field pulse is applied to the magnetic layer and the natural precession of the material is used to aid the switching.  These methods have allowed ``ringing-free'' switching at speeds an order of magnitude faster than the time constant of natural damping in the material.  For obvious reasons these developments do not help in sensor applications, in which case one seldom has much control over the field being sensed, neither its magnitude nor direction.

An alternative approach in contending undesirable magnetic precession is to increase the magnetization damping in the material.  This helps the magnetic moment reach equilibrium sooner, thus reducing the risk of thermal excitation to a different equilibrium position.  Increased damping can be achieved either by modifying the bulk of the material (e.g.~by alloying\cite{BaileyIEEETM01, SnorriPAT02}, ion irradiation\footnote{Argon ion irradiation was found to affect both magnetization and anisotropy quite significantly\cite{WoodsAPL02}.  It was also found to affect $\alpha$, although not very strongly compared with the effect of alloying. Unfortunately the results were found to be somewhat unpredictable.  Due to obvious difficulties with integration of irradiation with the device fabrication processes this method was abandoned.}, or ion implantation), or in the case of thin enough films by modifying their surface (e.g.~by introducing certain materials on the surface or by increasing the surface roughness\cite{MizukamiJJAP01, SnorriPRB02}).  

We took the latter approach, i.e.~to increase damping, and present results on the effect of alloying Permalloy, Ni$_{80}$Fe$_{20}$ (``Py'') with six different 4d and 5d transition metals (4d: Nb, Ru, Rh; 5d: Ta, Os, Pt) keeping the Ni to Fe ratio fixed.  We examined the effects these have on the static and dynamic magnetic properties of the materials.  Thus we hoped to find a way to increase damping in a manner that appears should lend itself to a greater variety of applications than surface treatment does, and that is generally effective for a wide range of film thickness (or sample size).

We deposited thin films (50~nm thick) of the ternary alloys by dc-magnetron
sputtering from ready-made targets with the desired composition.  The diluent concentration in the sputter targets was 6\%, except for Pt 10\% and Rh 5\%.  The alloy was deposited on top of a buffer layer of 4~nm of Ta
sitting on thermally oxidized Si-substrates.  Ta buffer layers are known to
promote small grains and (111) texture in Py\cite{Galtier94}. Finally, for
protection we used a capping layer of 4~nm thick Ta.  Rutherford backscattering
measurements on our 6\%~Os sample confirmed its composition to within 0.5\%.  
Depositions were carried out with an applied magnetic field in the plane of the
films, to encourage easy axis growth.  Static magnetic properties, i.e.\
coercive field, in-plane-anisotropy-field, and magnetization were measured with
a vibrating sample magnetometer (VSM). The magnetization damping properties
where extracted from results of a ferromagnetic resonance measurement by
assuming a Gilbert damping term in the Landau-Lifshitz equation,
\begin{equation}
    \label{eq:LLG}
    \frac{\dif\vec{M}}{\dif t} = -\gamma \vec{M} \crs \vec{H}_{\text{eff}} +
    \frac{\alpha}{M} \vec{M}\crs\frac{\dif \vec{M}}{\dif t}~~,
\end{equation}
and fitting the complex-valued susceptibility resonance to extract $\alpha$, the dimensionless Gilbert damping coefficient\cite{Brown78,SnorriPRB02}.  Here $\vec{M}$ is magnetization, $\gamma = g\left| e\right|/2mc$ is the gyromagnetic ratio and $\vec{H_{\text{eff}}}$ is the effective magnetic field seen by the magnetization, expressed in terms of the free energy as $\vec{H_{\text{eff}}} = -\nabla_{\vec{M}}\F$.

\begin{figure}[htb]
    \begin{center}
	\includegraphics[width=8.5cm]{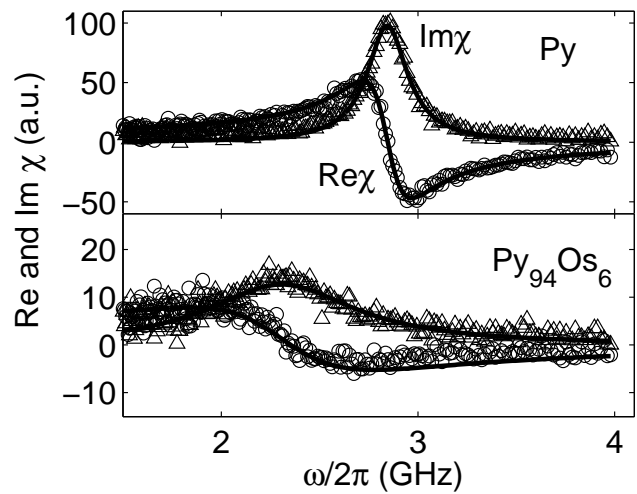}
    \end{center}
    \caption{Ingvarsson \emph{et al.}:  Results from a ferromagnetic resonance
    measurement on pure Py (top panel), and an alloy of 94\%~Py and 6\%~Os
    (bottom panel).  Each panel displays experimental data for both the real
    (circles) and the imaginary (triangles) part of the in-plane magnetic
    susceptibility to an ac-field perpendicular to the easy-axis direction with
    a 90~Oe dc-field applied along the easy axis.  Also shown as solid lines,
    are fits to the data, from which the Gilbert damping ($\alpha$) is
    obtained.}
    \label{fig:resonance}
\end{figure}

\begin{figure}[htb]
    \begin{center}
	\includegraphics[width=8.5cm]{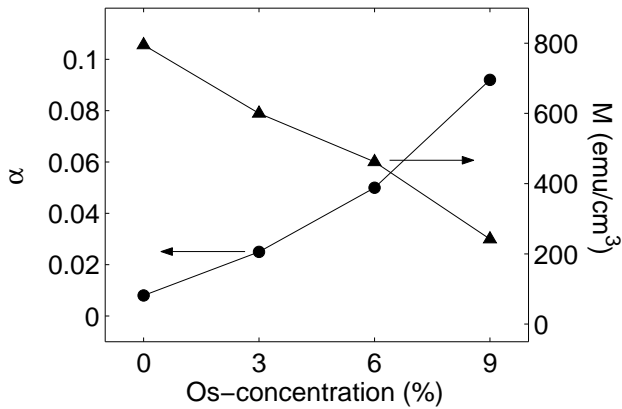}
    \end{center}
    \caption{Ingvarsson \emph{et al.}:  The effect of Os-concentration on both
    Gilbert damping $\alpha$ (left axis, dots), and magnetization $M$ (right
    axis, filled triangles).  The lines are guides to the eye.}
    \label{fig:concentration}
\end{figure}

When magnetization damping increases, i.e.\ $\alpha$ increases in Eq.~(\ref{eq:LLG}), the ferromagnetic resonance broadens and decreases in amplitude as is the case in Fig.~\ref{fig:resonance}.  
The top panel shows the real and imaginary parts of the complex-valued
susceptibility of pure Py around its resonance frequency.  The bottom panel
displays the corresponding data for Py containing 6\% of Os (osmium).  The
effect of adding Os was so dramatic that the vertical scale in the bottom panel
has been blown up and spans only a third of the scale in the top panel.  These
data correspond to more than a sixfold increase in $\alpha$ or $\alpha=0.050$.
The horizontal shift in the resonance towards lower frequency is caused
primarily by a decrease in magnetization upon adding Os.  The magnetization was
measured separately by VSM and included in the fitting procedure.  The level of
damping obtained goes a long way towards eliminating unwanted magnetization
precession.  However, in order to convince ourselves that adding Os to Py allows
sufficient tunability of the damping we made addition samples with Os-concentrations of
3\% and 9\%, respectively.  In Fig.~\ref{fig:concentration} the resulting Gilbert damping (left axis, dots) and
magnetization values (right axis, triangles) are presented.
The Gilbert damping increases and the magnetization decreases, respectively,
with increasing Os-concentration.  The value of $\alpha$ at 9\% Os is 0.092.
This corresponds to more than a 12-fold increase compared with pure Py.  Raising
$\alpha$ any further than this is likely to result in an overdamped system and
an unnecessary slowing down in its dynamic response.  In other words, adjustment
of the amount of Os in Py within the range shown should \emph{provide sufficient tunability of the
magnetization damping to minimize post-switching magnetization precession}.

\begin{figure}[htb]
    \begin{center}
	\includegraphics[width=8.5cm]{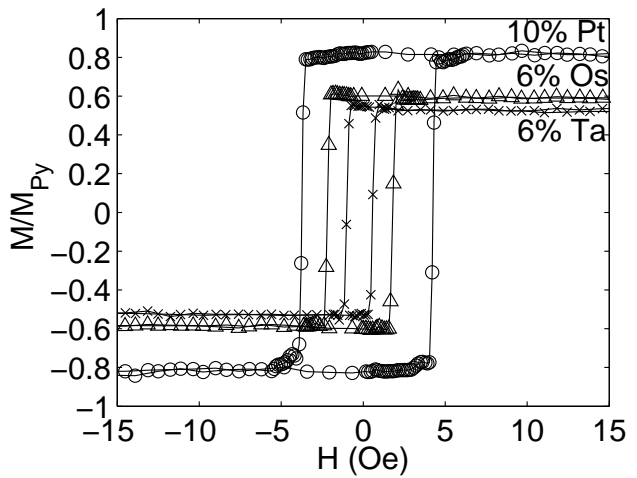}
    \end{center}
    \caption{Ingvarsson \emph{et al.}:  Hysteresis loops for Py containing
    10\%~Pt, 6\%~Os, and 6\%~Ta, respectively.  $H$ is the magnetic field
    applied parallel to the easy axis.  The magnetization is normalized to the
    saturation magnetization of pure Py.}
    \label{fig:5dloops}
\end{figure}

Although some of the other alloys exhibit significant enhancements in
$\alpha$ over the value for pure Py, none are comparable with the Py-Os
system.  Results for all the alloys are displayed in
Table~\ref{table:alphadoped}.
\begin{table}[htb]
    \caption{\label{table:alphadoped}Summary of results of Gilbert damping
    ($\alpha$) and static magnetic properties for alloys of Py$_{1-x}$X$_x$,
    where Py is Ni$_{80}$Fe$_{20}$ and X is a 4d or 5d transition metal.}
    \begin{ruledtabular}
    \begin{tabular}{cccccc}
	X & $\alpha$ $\left(\times 10^{-3}\right)$ & 
	$x$~(\%)\footnote{This is the concentration in the sputter target.  We
	did a composition study on the Os-sample, using Rutherford backscattering,
	and found that the composition of the film was the same as the target
	composition to within 0.5~\%.}
	& $M$~$(\frac{\text{emu}}{\text{cm}^3})$ & $H_{k}$~(Oe) & $H_{c}$~(Oe) \\\hline
	\multicolumn{6}{c}{Pure Permalloy (Ni$_{80}$Fe$_{20}$)}\\
	 & 7 &  & 796 & 5.1 & 1.0 \\
%
	\multicolumn{6}{c}{4d transition metal elements}\\
	Nb & 14 & 6 & 492 & 3.0 & 1.6\\
	Ru & 18 & 6 & 498 & 2.7 & 0.9\\
	Rh & 12 & 5 & 670 & 3.9 & 1.6\\
%
	\multicolumn{6}{c}{5d transition metal elements}\\
	Ta & 16 & 6 & 417 & 2.7 & 0.8\\
	Os & 50 & 6 & 462 & 4.6 & 2.0\\
	Pt & 20 & 10 & 649 & 3.0 & 4.0\\
    \end{tabular}
    \end{ruledtabular}
\end{table}
Results of VSM-measurements to determine static magnetic properties such as
magnetization, in-plane-anisotropy-field ($H_k$), and coercivity ($H_c$) are
also included.  Note that the VSM-measurements were made on macroscopic films
(i.e.\ unpatterned).  Therefore there is effectively no shape-anisotropy in the
plane of the film, and the in-plane anisotropy is a measure of average
crystalline anisotropy.  All the samples have a uniaxial magnetic anisotropy and
nice rectangular easy-axis hysteresis loops, some of which are displayed in
Fig.~\ref{fig:5dloops}.
From our data there does not appear to be any correlation between the static
magnetic properties and $\alpha$.  It seems that Rh and in particular Pt are
very ``polarizable'' in the sense that their addition to Py does not drop the
magnetization as much as the other elements do.  The anisotropy is enhanced by
Os but it is unclear how this should increase $\alpha$.

Obviously each of the properties listed in Table~\ref{table:alphadoped} can be
extremely important in practical applications.  If we consider in particular the
MRAM application, a decrease in magnetization associated with alloying Py with a
third element may be beneficial.  This would reduce dipolar coupling between
non-adjacent layers in the stack of layers needed to ensure proper
functionality.  The larger the crystalline anisotropy the better, as this
increases thermal stability of memory bits.  Also a low coercive field is
desirable, as coercivity tends to increase as bits are made smaller and heat
generation caused by large switching currents (to generate large switching
fields) is of great concern.  All the above considerations seem to favor the
addition of e.g.\ Os.  However, a most pressing question regarding the use of
any of the alloys studied in MRAM elements is:  What effect does the alloying
have on the spin polarization?  Although the levels of spin polarization
achieved to date are much larger than needed for MRAM to function well, a
significant reduction in polarization may cause problems\cite{ParkinJAP99}.
Therefore it is of substantial interest to study the spin polarization in these
materials.

In summary we have discovered that alloying Permalloy with certain transition
metal elements can enhance Gilbert damping significantly.  This should help
reduce, or even eliminate, superfluous magnetization precession that can
adversely affect or hinder many magnetoelectronic applications.

The authors would like to thank K.~Pope for the Rutherford backscattering
analysis, W.~J.~Gallagher, T.~M.~McGuire, and P.~L.~Trouilloud for many helpful
discussions. 


\begin{thebibliography}{17}
\expandafter\ifx\csname natexlab\endcsname\relax\def\natexlab#1{#1}\fi
\expandafter\ifx\csname bibnamefont\endcsname\relax
  \def\bibnamefont#1{#1}\fi
\expandafter\ifx\csname bibfnamefont\endcsname\relax
  \def\bibfnamefont#1{#1}\fi
\expandafter\ifx\csname citenamefont\endcsname\relax
  \def\citenamefont#1{#1}\fi
\expandafter\ifx\csname url\endcsname\relax
  \def\url#1{\texttt{#1}}\fi
\expandafter\ifx\csname urlprefix\endcsname\relax\def\urlprefix{URL }\fi
\providecommand{\bibinfo}[2]{#2}
\providecommand{\eprint}[2][]{\url{#2}}

\bibitem[{\citenamefont{Parkin et~al.}(1999)\citenamefont{Parkin, Roche,
  Samant, Rice, Beyers, Scheuerlein, {O'Sullivan}, Brown, Bucchigano, Abraham
  et~al.}}]{ParkinJAP99}
\bibinfo{author}{\bibfnamefont{S.~S.~P.} \bibnamefont{Parkin}},
  \bibinfo{author}{\bibfnamefont{K.~P.} \bibnamefont{Roche}},
  \bibinfo{author}{\bibfnamefont{M.~G.} \bibnamefont{Samant}},
  \bibinfo{author}{\bibfnamefont{P.~M.} \bibnamefont{Rice}},
  \bibinfo{author}{\bibfnamefont{R.~B.} \bibnamefont{Beyers}},
  \bibinfo{author}{\bibfnamefont{R.~E.} \bibnamefont{Scheuerlein}},
  \bibinfo{author}{\bibfnamefont{E.~J.} \bibnamefont{{O'Sullivan}}},
  \bibinfo{author}{\bibfnamefont{S.~L.} \bibnamefont{Brown}},
  \bibinfo{author}{\bibfnamefont{J.}~\bibnamefont{Bucchigano}},
  \bibinfo{author}{\bibfnamefont{D.~W.} \bibnamefont{Abraham}},
  \bibnamefont{et~al.}, \bibinfo{journal}{J. Appl. Phys.}
  \textbf{\bibinfo{volume}{85}}, \bibinfo{pages}{5828} (\bibinfo{year}{1999}).

\bibitem[{\citenamefont{Reohr et~al.}(2002)\citenamefont{Reohr, Honigschmid,
  Robertazzi, Gogl, Pesavento, Lammers, Lewis, Arndt, Lu, Viehmann
  et~al.}}]{Reohri3ecirc02}
\bibinfo{author}{\bibfnamefont{W.}~\bibnamefont{Reohr}},
  \bibinfo{author}{\bibfnamefont{H.}~\bibnamefont{Honigschmid}},
  \bibinfo{author}{\bibfnamefont{R.}~\bibnamefont{Robertazzi}},
  \bibinfo{author}{\bibfnamefont{D.}~\bibnamefont{Gogl}},
  \bibinfo{author}{\bibfnamefont{F.}~\bibnamefont{Pesavento}},
  \bibinfo{author}{\bibfnamefont{S.}~\bibnamefont{Lammers}},
  \bibinfo{author}{\bibfnamefont{K.}~\bibnamefont{Lewis}},
  \bibinfo{author}{\bibfnamefont{C.}~\bibnamefont{Arndt}},
  \bibinfo{author}{\bibfnamefont{Y.}~\bibnamefont{Lu}},
  \bibinfo{author}{\bibfnamefont{H.}~\bibnamefont{Viehmann}},
  \bibnamefont{et~al.}, \bibinfo{journal}{IEEE Circuits and Devices Magazine}
  \textbf{\bibinfo{volume}{18}}, \bibinfo{pages}{17} (\bibinfo{year}{2002}).

\bibitem[{\citenamefont{Liu and Xiao}(2003)}]{LiuJAP03}
\bibinfo{author}{\bibfnamefont{X.}~\bibnamefont{Liu}} \bibnamefont{and}
  \bibinfo{author}{\bibfnamefont{G.}~\bibnamefont{Xiao}}, \bibinfo{journal}{J.
  Appl. Phys.} \textbf{\bibinfo{volume}{94}}, \bibinfo{pages}{6218}
  (\bibinfo{year}{2003}).

\bibitem[{\citenamefont{Tondra et~al.}(2000)\citenamefont{Tondra, Daughton,
  Nordman, Wang, and Taylor}}]{TondraJAP00}
\bibinfo{author}{\bibfnamefont{M.}~\bibnamefont{Tondra}},
  \bibinfo{author}{\bibfnamefont{J.~M.} \bibnamefont{Daughton}},
  \bibinfo{author}{\bibfnamefont{C.}~\bibnamefont{Nordman}},
  \bibinfo{author}{\bibfnamefont{D.}~\bibnamefont{Wang}}, \bibnamefont{and}
  \bibinfo{author}{\bibfnamefont{J.}~\bibnamefont{Taylor}},
  \bibinfo{journal}{J. Appl. Phys.} \textbf{\bibinfo{volume}{87}},
  \bibinfo{pages}{4679} (\bibinfo{year}{2000}).

\bibitem[{\citenamefont{Schrag and Xiao}(2003)}]{SchragAPL03}
\bibinfo{author}{\bibfnamefont{B.~D.} \bibnamefont{Schrag}} \bibnamefont{and}
  \bibinfo{author}{\bibfnamefont{G.}~\bibnamefont{Xiao}},
  \bibinfo{journal}{Appl. Phys. Lett.} \textbf{\bibinfo{volume}{82}},
  \bibinfo{pages}{3272} (\bibinfo{year}{2003}).

\bibitem[{\citenamefont{Dieny}(1994)}]{DienyJMMM94}
\bibinfo{author}{\bibfnamefont{B.}~\bibnamefont{Dieny}}, \bibinfo{journal}{J.
  Magn. Magn. Mater.} \textbf{\bibinfo{volume}{136}}, \bibinfo{pages}{335}
  (\bibinfo{year}{1994}).

\bibitem[{\citenamefont{Koch et~al.}(1998)\citenamefont{Koch, Deak, Abraham,
  Troullioud, Altman, Lu, Gallagher, Scheuerlein, Roche, and
  Parkin}}]{KochPRL98}
\bibinfo{author}{\bibfnamefont{R.~H.} \bibnamefont{Koch}},
  \bibinfo{author}{\bibfnamefont{J.~G.} \bibnamefont{Deak}},
  \bibinfo{author}{\bibfnamefont{D.~W.} \bibnamefont{Abraham}},
  \bibinfo{author}{\bibfnamefont{P.~L.} \bibnamefont{Troullioud}},
  \bibinfo{author}{\bibfnamefont{R.~A.} \bibnamefont{Altman}},
  \bibinfo{author}{\bibfnamefont{Y.}~\bibnamefont{Lu}},
  \bibinfo{author}{\bibfnamefont{W.~J.} \bibnamefont{Gallagher}},
  \bibinfo{author}{\bibfnamefont{R.~E.} \bibnamefont{Scheuerlein}},
  \bibinfo{author}{\bibfnamefont{K.~P.} \bibnamefont{Roche}}, \bibnamefont{and}
  \bibinfo{author}{\bibfnamefont{S.~S.~P.} \bibnamefont{Parkin}},
  \bibinfo{journal}{Phys. Rev. Lett.} \textbf{\bibinfo{volume}{81}},
  \bibinfo{pages}{4512} (\bibinfo{year}{1998}).

\bibitem[{\citenamefont{Crawford et~al.}(2000)\citenamefont{Crawford, Kabos,
  and Silva}}]{CrawfordAPL00}
\bibinfo{author}{\bibfnamefont{T.~M.} \bibnamefont{Crawford}},
  \bibinfo{author}{\bibfnamefont{P.}~\bibnamefont{Kabos}}, \bibnamefont{and}
  \bibinfo{author}{\bibfnamefont{T.~J.} \bibnamefont{Silva}},
  \bibinfo{journal}{Appl. Phys. Lett.} \textbf{\bibinfo{volume}{76}},
  \bibinfo{pages}{2113} (\bibinfo{year}{2000}).

\bibitem[{\citenamefont{Gerrits et~al.}(2002)\citenamefont{Gerrits, {van den
  Berg}, Hohlfeld, B{\"{a}}r, and Rasing}}]{GerritsNAT02}
\bibinfo{author}{\bibfnamefont{T.}~\bibnamefont{Gerrits}},
  \bibinfo{author}{\bibfnamefont{H.~A.~M.} \bibnamefont{{van den Berg}}},
  \bibinfo{author}{\bibfnamefont{J.}~\bibnamefont{Hohlfeld}},
  \bibinfo{author}{\bibfnamefont{L.}~\bibnamefont{B{\"{a}}r}},
  \bibnamefont{and} \bibinfo{author}{\bibfnamefont{T.}~\bibnamefont{Rasing}},
  \bibinfo{journal}{Nature} \textbf{\bibinfo{volume}{418}},
  \bibinfo{pages}{509} (\bibinfo{year}{2002}).

\bibitem[{\citenamefont{Schumacher et~al.}(2003)\citenamefont{Schumacher,
  Chappert, Crozat, Sousa, Freitas, Miltat, Fassbender, and
  Hillebrands}}]{SchumacherPRL03b}
\bibinfo{author}{\bibfnamefont{H.~W.} \bibnamefont{Schumacher}},
  \bibinfo{author}{\bibfnamefont{C.}~\bibnamefont{Chappert}},
  \bibinfo{author}{\bibfnamefont{P.}~\bibnamefont{Crozat}},
  \bibinfo{author}{\bibfnamefont{R.~C.} \bibnamefont{Sousa}},
  \bibinfo{author}{\bibfnamefont{P.~P.} \bibnamefont{Freitas}},
  \bibinfo{author}{\bibfnamefont{J.}~\bibnamefont{Miltat}},
  \bibinfo{author}{\bibfnamefont{J.}~\bibnamefont{Fassbender}},
  \bibnamefont{and}
  \bibinfo{author}{\bibfnamefont{B.}~\bibnamefont{Hillebrands}},
  \bibinfo{journal}{Phys. Rev. Lett.} \textbf{\bibinfo{volume}{90}},
  \bibinfo{pages}{017201} (\bibinfo{year}{2003}).

\bibitem[{\citenamefont{Bailey et~al.}(2001)\citenamefont{Bailey, Kabos,
  Mancoff, and Russek}}]{BaileyIEEETM01}
\bibinfo{author}{\bibfnamefont{W.}~\bibnamefont{Bailey}},
  \bibinfo{author}{\bibfnamefont{P.}~\bibnamefont{Kabos}},
  \bibinfo{author}{\bibfnamefont{F.}~\bibnamefont{Mancoff}}, \bibnamefont{and}
  \bibinfo{author}{\bibfnamefont{S.}~\bibnamefont{Russek}},
  \bibinfo{journal}{IEEE Trans. Magn.} \textbf{\bibinfo{volume}{37}},
  \bibinfo{pages}{1749} (\bibinfo{year}{2001}).

\bibitem[{\citenamefont{Ingvarsson et~al.}()\citenamefont{Ingvarsson, Koch,
  Parkin, and Xiao}}]{SnorriPAT02}
\bibinfo{author}{\bibfnamefont{S.}~\bibnamefont{Ingvarsson}},
  \bibinfo{author}{\bibfnamefont{R.}~\bibnamefont{Koch}},
  \bibinfo{author}{\bibfnamefont{S.~S.~P.} \bibnamefont{Parkin}},
  \bibnamefont{and} \bibinfo{author}{\bibfnamefont{G.}~\bibnamefont{Xiao}},
  \bibinfo{journal}{U.S.~Patent no.~6,452,240}, \bibinfo{note}{filed
  Oct.~30 2000, issued Sept.~17 2002}.

\bibitem[{\citenamefont{Mizukami et~al.}(2001)\citenamefont{Mizukami, Ando, and
  Miyazaki}}]{MizukamiJJAP01}
\bibinfo{author}{\bibfnamefont{S.}~\bibnamefont{Mizukami}},
  \bibinfo{author}{\bibfnamefont{Y.}~\bibnamefont{Ando}}, \bibnamefont{and}
  \bibinfo{author}{\bibfnamefont{T.}~\bibnamefont{Miyazaki}},
  \bibinfo{journal}{Jpn. J. Appl. Phys.} \textbf{\bibinfo{volume}{40}},
  \bibinfo{pages}{580} (\bibinfo{year}{2001}).

\bibitem[{\citenamefont{Ingvarsson et~al.}(2002)\citenamefont{Ingvarsson,
  Ritchie, Liu, Xiao, Slonczewski, Trouilloud, and Koch}}]{SnorriPRB02}
\bibinfo{author}{\bibfnamefont{S.}~\bibnamefont{Ingvarsson}},
  \bibinfo{author}{\bibfnamefont{L.}~\bibnamefont{Ritchie}},
  \bibinfo{author}{\bibfnamefont{X.~Y.} \bibnamefont{Liu}},
  \bibinfo{author}{\bibfnamefont{G.}~\bibnamefont{Xiao}},
  \bibinfo{author}{\bibfnamefont{J.~C.} \bibnamefont{Slonczewski}},
  \bibinfo{author}{\bibfnamefont{P.~L.} \bibnamefont{Trouilloud}},
  \bibnamefont{and} \bibinfo{author}{\bibfnamefont{R.~H.} \bibnamefont{Koch}},
  \bibinfo{journal}{Phys. Rev. B} \textbf{\bibinfo{volume}{66}},
  \bibinfo{pages}{214416} (\bibinfo{year}{2002}), \bibinfo{note}{{NB}. There is
  an awkward typographical error in Eq.~(1). The sign of the second term on the
  right hand side should be reversed from $-$ to $+$. This affects neither the
  analysis nor the results in the paper.}

\bibitem[{\citenamefont{Galtier et~al.}(1994)\citenamefont{Galtier, Jerome, and
  Valet}}]{Galtier94}
\bibinfo{author}{\bibfnamefont{P.}~\bibnamefont{Galtier}},
  \bibinfo{author}{\bibfnamefont{R.}~\bibnamefont{Jerome}}, \bibnamefont{and}
  \bibinfo{author}{\bibfnamefont{T.}~\bibnamefont{Valet}}, in
  \emph{\bibinfo{booktitle}{Polycrystalline thin films: Structure, texture,
  properties and applications}} (\bibinfo{publisher}{Mater. Res. Soc.},
  \bibinfo{address}{Pittsburgh}, \bibinfo{year}{1994}), pp.
  \bibinfo{pages}{417--422}.

\bibitem[{\citenamefont{Brown}(1978)}]{Brown78}
\bibinfo{author}{\bibfnamefont{W.~F.} \bibnamefont{Brown}},
  \emph{\bibinfo{title}{Micromagnetics}} (\bibinfo{publisher}{Krieger
  Publishing Company}, \bibinfo{address}{Melbourne, Florida, U.S.A.},
  \bibinfo{year}{1978}).

\bibitem[{\citenamefont{Woods et~al.}(2002)\citenamefont{Woods, Ingvarsson,
  Kirtley, Hamann, and Koch}}]{WoodsAPL02}
\bibinfo{author}{\bibfnamefont{S.~I.} \bibnamefont{Woods}},
  \bibinfo{author}{\bibfnamefont{S.}~\bibnamefont{Ingvarsson}},
  \bibinfo{author}{\bibfnamefont{J.~R.} \bibnamefont{Kirtley}},
  \bibinfo{author}{\bibfnamefont{H.~F.} \bibnamefont{Hamann}},
  \bibnamefont{and} \bibinfo{author}{\bibfnamefont{R.~H.} \bibnamefont{Koch}},
  \bibinfo{journal}{Appl. Phys. Lett.} \textbf{\bibinfo{volume}{81}},
  \bibinfo{pages}{1267} (\bibinfo{year}{2002}).

\end{thebibliography}

\end{document}